\documentclass[twocolumn,floatfix,pra,aps,showpacs,superscriptaddress]{revtex4}
\usepackage{epsfig,graphicx}
\usepackage{flafter}
\usepackage{amsmath}
\usepackage{color}
\usepackage{bm}
\usepackage{amssymb}
\usepackage{epstopdf}
\usepackage{amsmath}
\usepackage{amstext}
\usepackage{amsthm}
\usepackage{amsfonts}
\usepackage{latexsym}
\usepackage{multirow}

\newcommand{\da}{^\dagger}
\newcommand{\ket}[1]{\left\vert#1\right\rangle}
\newcommand{\bra}[1]{\left\langle#1\right\vert}

\renewcommand{\emph}[1]{{\it #1}}
\renewcommand{\vec}[1]{\boldsymbol{#1}}

\begin{document}

\title{Quantum Magnetism with Mesoscopic Bose-Einstein Condensates} 

\author{A. Gallem\'{\i}}
\affiliation{Departament de F\'{\i}sica Qu\`antica i Astrof\'{\i}sica, %Facultat de F\'{\i}sica, 
Universitat de Barcelona, E--08028
Barcelona, Spain}
\affiliation{Institut de Nanoci\`encia i Nanotecnologia de la Universitat de Barcelona, IN$\,^2$UB, E--08028 Barcelona, Spain}
\author{G. Queralt\'o}
\affiliation{Departament de F\'{\i}sica, Universitat
Aut\`{o}noma de Barcelona, E--08193 Bellaterra, Spain}
\author{M. Guilleumas}
\affiliation{Departament de F\'{\i}sica Qu\`antica i Astrof\'{\i}sica, %Facultat de F\'{\i}sica, 
Universitat de Barcelona, E--08028
Barcelona, Spain}
\affiliation{Institut de Nanoci\`encia i Nanotecnologia de la Universitat de Barcelona, IN$\,^2$UB, E--08028 Barcelona, Spain}
\author{R. Mayol}
\affiliation{Departament de F\'{\i}sica Qu\`antica i Astrof\'{\i}sica, %Facultat de F\'{\i}sica, 
Universitat de Barcelona, E--08028
Barcelona, Spain}
\affiliation{Institut de Nanoci\`encia i Nanotecnologia de la Universitat de Barcelona, IN$\,^2$UB, E--08028 Barcelona, Spain}
\author{A. Sanpera}
\affiliation{Departament de F\'{\i}sica, Universitat
Aut\`{o}noma de Barcelona, E--08193 Bellaterra, Spain}
\affiliation{Instituci\'o Catalana de Recerca i Estudis Avan\c{c}ats, ICREA, E--08011 Barcelona, Spain}

\date{\today}

\begin{abstract}

Lattice gases in the strongly correlated regime have been proven to simulate quantum magnetic models under 
certain conditions: the mapping of the double-well system onto the Lipkin-Meshkov-Glick spin model is a 
paradigmatic case. A suitable definition of the length in the Hilbert space of the system leads to the 
concept of a correlation length, whose divergence is a characteristic property of continuous quantum phase 
transitions. 
% 
% Here we show that mesoscopic Bose-Einstein condensates can also simulate quantum magnetic systems, a 
% paradigmatic case being the mapping of the double-well condensate onto the Lipkin-Meshkov-Glick spin model. 
% Furthermore, it is possible to define for such models a correlation length, and describe its continuous 
% quantum phase transitions by means of the divergence of a correlation length
% 
% 
% Lattice gases in the strongly correlated regime have been proven to simulate spin models under certain 
% conditions. Here we show that mesoscopic Bose gases trapped in few traps can be also mapped onto 
% spin systems, like the Lipkin-Meshkov-Glick model for the double-well case. 
% A suitable definition of the length in the corresponding Hilbert space of the system leads to the concept of 
% correlation length, whose divergence is a characteristic property of continuous quantum phase transitions. 
% 
% 
We calculate the finite-size scaling of some observables like e.g. the magnetization or the population 
imbalance, as well as of the Schmidt gap, obtaining in this way  the critical exponents associated to such 
transitions. 
%A suitable definition of length 
%in the Hilbert space of these systems can lead to the appearance of a correlation length, whose divergence 
%is a characteristic property in the vicinity of a quantum phase transition. Finite-size scaling of some 
%observables and the Schmidt gap gives the critical exponents associated to such transitions. 
The systematic 
definition of the Schmidt gap in extended Hamiltonians provides a good tool to analyze the set of critical 
exponents associated to transitions in systems formed by a larger number of traps. 
This demonstrates, thus, the potential use of mesoscopic 
Bose-Einstein condensates as quantum simulators of condensed matter systems.

\end{abstract}


\maketitle

\section{Introduction}

The achievement of quantum magnetism with ultracold lattice atoms can be regarded as one of the cornerstones of modern physics. 
Indeed, the seminal proposal \cite{Jaksch1998} that ultracold atoms confined in optical lattices can be described by a Bose-Hubbard 
model \cite{Fisher1989}, together with the experimental 
signatures of the superfluid-Mott insulator quantum phase transition in such systems \cite{Greiner2002} have prompted the field of 
quantum simulators. Since then,  the scope of condensed matter phenomena that can be addressed with ultracold gases and ions 
has broadened enormously (see e.g \cite{Bloch2008,Lewenstein2007,Lewenstein2012,Greif2013} and references therein).

One of the paradigmatic models of strongly correlated states in condensed matter are the family of Hubbard  Hamiltonians. 
These Hamiltonians describe, in a very simplified way, magnetic properties of some materials. 
Ultracold atoms loaded in optical lattices can realize, almost perfectly, many of such Hamiltonians. For instance, the Mott 
insulator phase of 
the Bose-Hubbard model is achieved by ``freezing'' at each lattice site a single atom. In certain limits, 
the Hubbard Hamiltonians reduce to various spin models: an exemplary case is the non-interacting 
XY spin model, which has been simulated with hard-core bosons in optical lattices \cite{Simon2011}. 
A generic constraint to realize interacting 
spin models with ultracold lattice atoms is the extreme low temperature required, going down to the picoKelvin regime.
The reason is that these spin models are derived as second order perturbation theory from Hubbard models, and temperature scales 
then as $t^{2}/U$ \cite{Staudt2000}, where $t$ denotes tunneling between nearest sites and $U$ is the atomic two-body onsite interaction.

Here, we take a different path to show that quantum magnetism can also be approached using mesoscopic (dipolar) 
Bose-Einstein condensates constituted by few cents of atoms confined in few (for the sake of simplicity we 
restrict to two and three) harmonic traps. In these systems, the relevant 
temperature is the critical temperature for degeneracy, which in the double-well potential is of the order of tens or 
hundreds of nanoKelvins, relaxing substantially the temperature constraints required by atoms in optical lattices to 
simulate spin models. Our main result is to show that these systems can be used to simulate quantum magnetism and, in 
particular, quantum phase transitions.

Our manuscript is organized as follows. In Section II we review first the mean-field description of a condensate in a 
double-well potential, and the regime for which the mean-field description fails due to the presence of a quantum phase 
transition. In Section III we analyze the quantum phase transition from a quantum magnetism perspective linking our results 
to key concepts like correlation length, critical exponents and universality, which are well defined in spin models, like 
the Ising model. We complement our study by analyzing these quantum phase transitions from an entanglement perspective. 
In Section IV, we extend our studies to dipolar condensates confined in three wells and show the meaning of universality 
classes in such systems. In Section V we present our conclusions and open questions.

\section{Mesoscopic condensates in a double-well potential}

The description of mesoscopic ultracold Bose-Einstein condensates confined in few-well potentials has been addressed 
in great detail both experimentally \cite{Morsch2001,Albiez2005,Levy2007} and theoretically, see e.g. \cite{Morsch2006} and 
references therein.  
In the weakly interacting regime, these mesoscopic systems can be described within a Hartree approach in which all particles 
share a common state (the condensate wavefunction) and the dynamics at low temperatures can then be accurately reproduced with 
the  time-dependent Gross-Pitaevskii equation. For a double-well potential, the description 
of the system can be further simplified with the so-called two-mode approximation, where the population imbalance 
and the phase difference between the condensates on each well are enough to capture the physics displayed by the system.  
Such mean field description, nevertheless, fails drastically when the interaction strength between atoms approaches some critical 
value. At this moment, the Gross-Pitaevskii solution breaks the symmetry of the double-well potential and becomes highly 
unstable \cite{JuliaDiaz2010a}. In these cases, the Bogoliubov approach 
shows a divergence in the number of atoms in the non-condensate modes \cite{Zin2008}. 
In such regime, an accurate description of the system is a simplified Bose-Hubbard Hamiltonian 
where each trap corresponds now to a mode or site.

In what follows we consider $N$ spinless bosons trapped in $M$ sites, 
which we restrict to be $M\le 3$, although our study could be generalized to higher $M$. We further assume that atoms might 
have an electric or magnetic dipole moment $\vec{d}$, 
all of them polarized along the same direction by the presence of an external strong field. Bosons interact 
via short range potentials but also, when present, with dipolar long-range interactions that 
couple bosons in different traps. The bosonic field operators that annihilate (create) a boson at a point $\vec{r}$ are defined as 
$\hat{\psi}(\vec{r})=\sum_{i} \phi_{i}(\vec{r}) \, \hat{a}_{i}$, where as usual $\hat{a}_{i} (\,\hat{a}_{i}^{\dagger})$ is the 
bosonic annihilation (creation) operator on trap $i$ fulfilling canonical commutation relations 
$[\hat{a}_i^{\dag},\hat{a}_j]=\delta_{ij}$. Under these assumptions, 
the extended Bose-Hubbard Hamiltonian reads:

\begin{align}
{\hat H}=&-\frac{t}{2}\sum_{i}\left[\hat{a}^\dagger_{i} \hat{a}_{i+1} + h.c. \right]
+\frac{U}{2} \sum_{i} \hat{n}_i (\hat{n}_i-1) \nonumber \\
&+\sum_{i\neq j} U_{ij} \, \hat{n}_i \hat{n}_j\,,
\label{BHH}
\end{align}
where $\hat{n}_{i}=\hat{a}_{i}^{\dagger}\hat{a}_{i}$ is the particle number operator on $i$-$th$ well, and 
$\sum_{i}\hat{n}_{i}=N$. 
In Eq. (\ref{BHH}), the indices $i,j$ run for all the lattice sites.
The Hamiltonian (\ref{BHH}) is 
characterized by three parameters: the tunneling rate $t$ between adjacent wells, the onsite energy $U$, which includes both 
contact and dipole-dipole interactions, and  the intersite energy $U_{ij}$, which takes into account the long range and anisotropy 
of the dipolar interaction. We notice here that for a double-well potential the effects of dipolar 
interactions can be included into a rescaled contact interaction, and the Hamiltonian reduces to the non-dipolar case 
\cite{Abad2011b}. The structure of the above Hamiltonian makes it convenient to work in the Fock basis 
$|F\rangle_{n_1, n_2,...,n_M}= |n_1,n_2,...,n_M \rangle = |n_1 \rangle\otimes |n_2 \rangle\otimes \cdots\otimes|n_M \rangle$ that labels the 
number of atoms in each well. Then, the wavefunction can be written as 
$\ket{\Psi} = \sum_{n_{i}} \, C_{n_{1},n_{2},...,n_{M}} \ket {F}_{n_{1},n_{2},...,n_{M}}$. The ground 
state solution is obtained by exact diagonalization of Eq. (\ref{BHH}), for different values of $N$, $t$ and $U$. 

For simplicity we 
consider first the double-well potential. In this case, the Schwinger representation allows to map the two-mode annihilation and 
creation operators onto spin operators:
\begin{eqnarray}
\hat{S}_{+}&=&\hat{a}_{1}^{\da} \hat{a}_{2},\nonumber\\
\hat{S}_{-}&=&\hat{a}_{2}^{\da} \hat{a}_{1}, \nonumber\\
\hat{S}_{z}&=&\frac{1}{2}(\hat{a}_{1}^{\da} \hat{a}_{1}-\hat{a}_{2}^{\da} \hat{a}_{2}),
\end{eqnarray}
with the constraint $\hat{a}_{1}^{\da}\hat{a}_{1}+\hat{a}_{2}^{\da} \hat{a}_{2}=2\hat{S}=\hat{N}$, which 
fixes the total number of bosons or equivalently the total spin.
Using such representation, the two-site Bose-Hubbard 
Hamiltonian can be rewritten as
\begin{eqnarray}
\hat{H}_{DW}&=&-\frac{t}{2} (\hat{S}_{+}+\hat{S}_{-}) + U (\hat{S}_{z}^{2} +\hat{S}^{2}-\hat{S})\nonumber\\
&=&-t\hat{S}_{x}+U \,\hat{S}_{z}^{2},
\label{BBHLMG}
\end{eqnarray}
where in the last equation, we have used that $[\hat{H}_{DW},\hat{N}]=[\hat{H}_{DW},\hat{S}]=0$ to 
remove all terms proportional to the total spin $\hat{S}$. 
By defining $\hat{S}_{\alpha}=\sum_{i=1}^{N}\hat{\sigma}_{i}^{\alpha}/2$ as the collective spin along the 
$\alpha$-direction, 
where the set of operators $\sigma^\alpha$ corresponds to the Pauli matrices, 
the double-well Hamiltonian $\hat{H}_{DW}$ can be interpreted as a system of $N$ spin-$1/2$ particles mutually 
interacting along the $z$-axis and embedded in a transverse magnetic field along the $x$-direction. 
Thus, the double-well Hamiltonian (\ref{BBHLMG}) is just a particular case of the general 
Lipkin-Meshkov-Glick model \cite{Lipkin1965} introduced long time ago in nuclear physics 
to study quantum phase transitions (QPTs) in ``mean field'' models, where all spins interact 
to each other, and since then exploited in many different contexts e.g. 
\cite{Botet1983,Reslen2005,Dusuel2004,Orus2008}. The corresponding Hamiltonian is:
\begin{equation}
\hat{H}_{LMG}=-h\sum_i\hat{\sigma}_{i}^{x}-\frac{\lambda}{N}\,\sum_{i<j}\left(\hat{\sigma}_{i}^{z}\hat{\sigma}_{j}^{z}+\gamma\hat{\sigma}_{i}^{y}\hat{\sigma}_{j}^{y}\right),
\end{equation}
where the factor $1/N$ ensures the convergence of the free energy per spin in the thermodynamic limit (TL) 
$N\rightarrow \infty$. This 
magnetic model and its corresponding phase transitions have been very well studied. For $\lambda>0$,  i.e. when the 
interaction between spins is ferromagnetic, there exists a second order 
phase transition at $\lambda=|h|$, if $0\leq\gamma\leq 1$. 
Thus, approaching  $|U|N/t  \rightarrow 1$, there is a QPT between 
ferromagnetic and paramagnetic order, which in the TL converges 
to $|U|/t \rightarrow 0$. The paramagnetic region is thus proportional to $1/N$, non-degenerate and 
has a gap that vanishes at criticality. 
We notice that in the double-well potential, the TL corresponds to the 
limit in which a mean field description is valid and does not require to reach $N\rightarrow \infty$ 
but rather a sufficiently large $N$. Finally, we remark that in these models, in which each particle 
interacts with each other, the dimension of the Hilbert space is infinite (mean field) and 
the concept of length is not defined. 

\section{Correlation length and critical exponents in a double-well potential}

As already mentioned, 
when approaching the limit $|U|N/t\rightarrow1$, the mean field 
description of the double well collapses and the Gross-Pitaevskii equation 
cannot provide anymore the description of the 
ground state of the system. The two-well Bose-Hubbard description shows also a massive fluctuation 
of the particle number on each well in the border of the transition regime. Indeed, it is the existence 
of quantum fluctuations on all length scales, the most characteristic feature of (continuous) QPTs occurring at 
zero-temperature. This behavior, commonly denoted as criticality, is also reflected on the behavior of some observables 
$\hat O$ that scale near the transition point $|U_{crit}|= t/N$ as a power law $ \hat O\propto |U-U_{crit}|^{\alpha}$, 
where $\alpha$ is a set of parameters called critical exponents that determine the qualitative nature of 
the critical behavior. Those parameters $\alpha$ are independent of the microscopic details of the system, 
but are rather linked to the symmetries of the emerging order as well as to the dimensionality of the 
system. Thus,  QPTs associated to different Hamiltonians that share the same set of critical exponents 
are said to belong to the same universality class. The QPT is also accompanied by the vanishing of some 
energy scale and the divergence of some length (the correlation length) which indicates the spread of 
correlations in the system \cite{Sachdev1999}.

We aim at interpreting the QPTs in the double-well potential in a similar way as in spin chains, 
provided a definition of correlations. This definition allows to link critical behavior to the 
divergence of a correlation length. 
With this purpose we first calculate the phase diagram of the double well for different values of $N$ near 
criticality. Inspired by two-point correlations in spin chains, we define correlations in our system 
in such a way that their behavior properly displays the relevant features of the QPTs.
Then, we analyze the scaling behavior of some operators and perform finite-size scaling (see below) to obtain the 
corresponding critical exponents. Finally, we check if other models of the restricted Bose-Hubbard family 
share the same critical exponents and belong, therefore, to the same universality class.

\begin{figure}[t!]
\centering
\epsfig{file=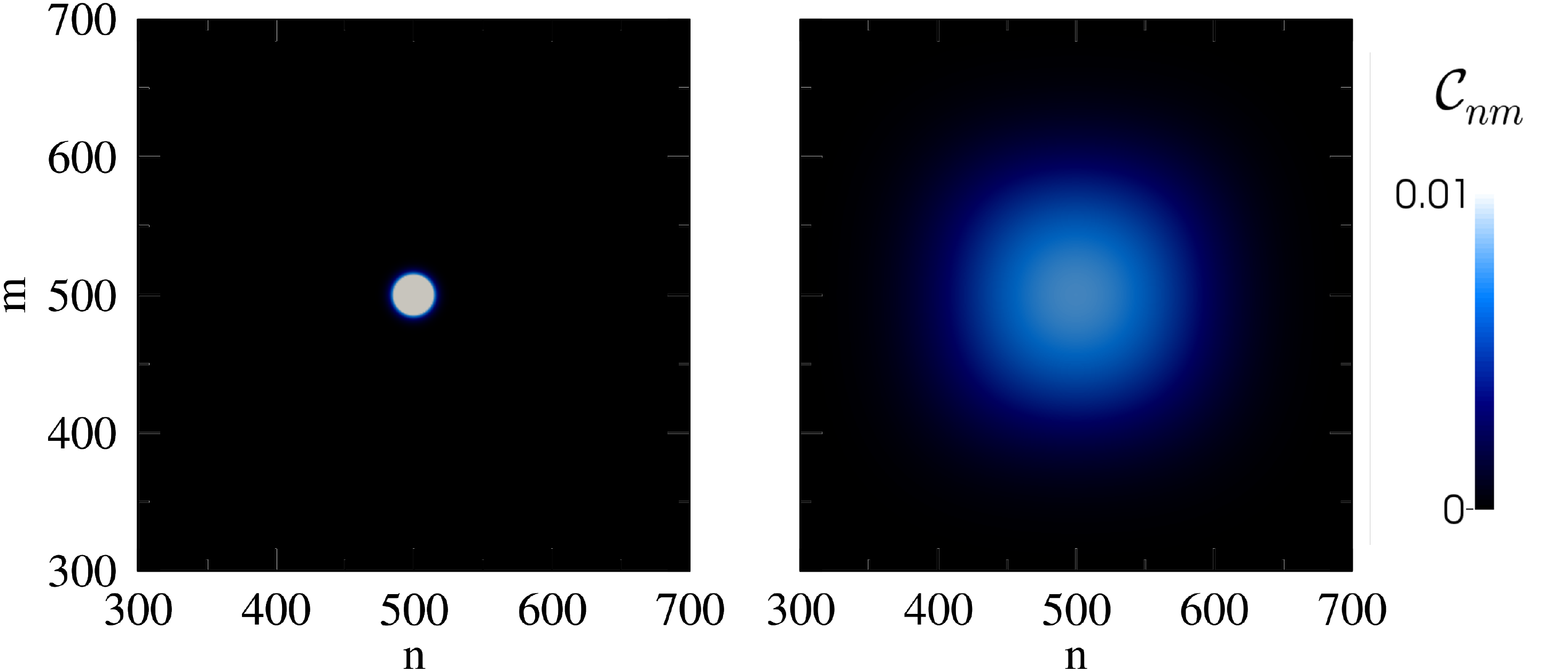, width=0.99\linewidth, clip=true}
\caption{Behavior of the correlations $\mathcal{C}_{nm}$ as a function of $n$ and $m$ for a double well. 
Left panel:  $UN/t=100$ far from criticality; right panel: $UN/t=-1$, at 
criticality. The figures correspond to $N=1000$, thus, the 
ground state corresponds to $500$ atoms in each well (in the center of color map).}
\label{Fig1}
\end{figure}

Notice that in a spin chain of spin $\vec{S}$ particles, the length is naturally settled by the 
number of sites $L$ (number of spins), and two-body spin-spin correlations are given by 
$\mathcal{C}_{ij}=\langle \vec{S}_{i}\vec{S}_{j}\rangle-\langle \vec{S}_{i}\rangle\langle\vec{S}_{j}\rangle$. 
Translational invariance in such systems ensures that the behavior of spin-spin correlations
depends only on the distance between the two sites $|i-j|$, but not on the specific sites $i,j$. 
This fact allows to define a correlation length $\xi$, which fixes the length scale at which 
the spins are correlated between them. Far from criticality, the  decay of correlations is 
exponential $\mathcal{C}_{ij}\sim \exp(-|i-j|/\xi)$, meaning that two far away spins are not 
correlated. At criticality, for continuous second order phase transitions, the decay is 
algebraic  $\mathcal{C}_{ij} \sim (|i-j|^{-(d-2+\eta)})$, where $d$ is the dimension, and the correlation length 
diverges as $\xi \propto |U-U_{crit}|^{-\nu}$, 
where $U_{crit}$ is the critical point, expressing the fact that now spins are correlated 
between them even when they are far.

\begin{figure}[t!]
\centering
\epsfig{file=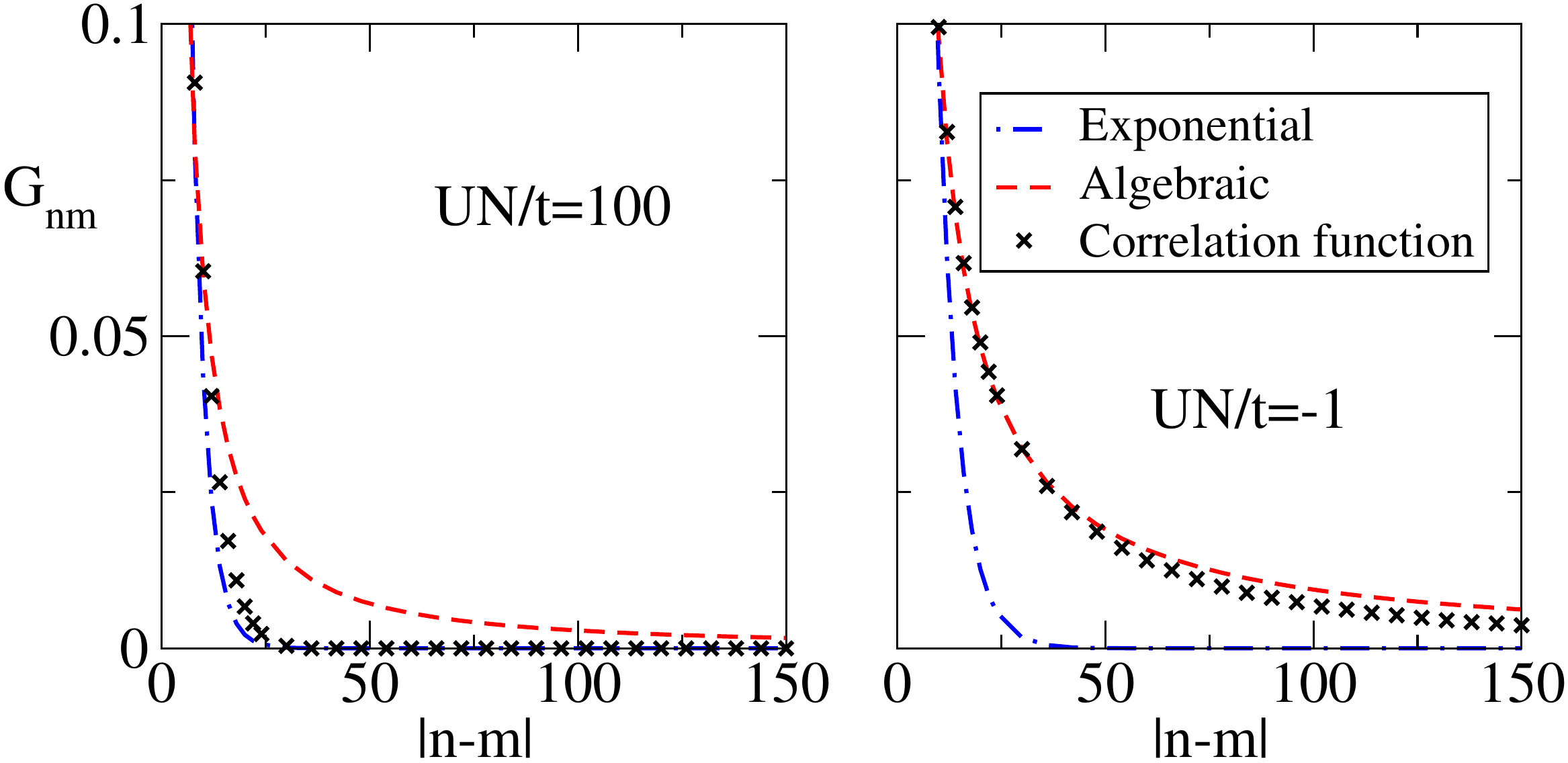, width=0.99\linewidth, clip=true}
\caption{Correlation function for the double-well potential 
$G_{nm}$ versus $|n-m|$. Left panel: $UN/t=100$ far 
from criticality. Right panel: $UN/t=-1$ at the critical point. 
The figures correspond to $N=1000$. To better understand the 
different behavior far and close to criticality, an exponential 
decay (dashed-dotted line) and an algebraic decay (dashed line) 
has been drawn to guide the eye.  As clearly displayed in the 
figure, at criticality (right panel) the best adjustment is algebraic, 
otherwise it is exponential (left panel).}
\label{Fig2}
\end{figure}

In a double well, such length scale is obviously irrelevant. In order to mimic the 
behavior of second order QPTs in 
spin chains in our system, we start by realizing that the quantity that settles the 
dimension of the Hilbert space here is the number of bosons on each well. Furthermore, 
this permits to order the Fock states 
in the following way: $|N,0\rangle,|N-1,1\rangle,\cdots,|0,N\rangle$. 
We then define two-body correlations in the ground state of our system as: 
\begin{eqnarray}
 G_{nm}&=&\sum_{|n-m|}\frac{\langle\,\ket{n}\bra{m} \otimes \ket{ N-n}\bra{N-m}\,\rangle}{|n-m|}\nonumber\\
 &=&\sum_{|n-m|} \frac{\mathcal{C}_{nm}}{|n-m|}\,,
 \label{correfunc}
\end{eqnarray}
where the operator $\ket{n}\bra{m}$ acts on the first trap and $\ket{ N-n}\bra{N-m}$ on the 
second one. So, we analyze how the number of bosons are correlated within a trap, i.e. 
the simultaneous presence of $n$ and $m$ bosons in the ground state and the sum extends 
over all possible contributions with $|n-m|$  fixed. 
Notice that $\mathcal{C}_{nm}$  is not equivalent to the widely used population imbalance 
operator in the double-well potential. The latter indicates the difference of population 
between the wells and corresponds to the expectation value of the population imbalance 
$Z=\langle \hat{S}_{z}\rangle=\langle |\hat{n}_{1}-\hat{n}_{2}|\rangle/N=\sum_{n_{1},n_{2}}|C_{n_{1},n_{2}}|^{2}(|n_{1}-n_{2}|)/N$, 
while the former, $\mathcal{C}_{nm}=C_{n,N-n}C_{m,N-m}$, correlates boson occupation 
numbers within a well. In order to recover the ``translational invariance'', concept 
rooted in spin chains, we weight the correlation function
by the ``effective distance'' $|n-m|$. This renormalization factor ensures the 
proper behavior of correlations as it neglects contributions from any intermediate 
level between $n$ and $m$.

The general behavior of the correlations $\mathcal{C}_{nm}$ as a function of $n$ and $m$, far from and close 
to criticality, is shown in Fig. \ref{Fig1}, in the left and right panel, respectively.
While far from criticality, the ground state is correlated with the closer states only, at criticality the spreading 
of correlations becomes evident. Nevertheless, it is indeed the weighted 
function $G_{nm}$ the one which allows to recover the concept of
``translational invariance''  and define 
a correlation length $\xi$ for the system. In Fig. \ref{Fig2}, we display $G_{nm}$ 
as a function of $|n-m|$ far from and near criticality. One can see that $G_{nm}$ decays 
exponentially far from the critical point and algebraically at criticality. To better 
stress the character of the correlation decay we fit the 
functions by an exponential and a power law.

To obtain a deeper understanding about the concept of correlation 
and the exact location of the QPT in the mean field limit, we analyze 
the behavior of the population imbalance $Z$, its fluctuations $\Delta Z$, 
the magnetization along the $z$-axis, defined as 
$m_z=(\langle \hat{S}_z^2 \rangle)^{1/2}/N$ \cite{Reslen2005}, and the entanglement spectrum. 
The latter is defined as 
the eigenvalues of the reduced density matrix of one mode or trap 
$\hat{\rho}_{L}=\mbox{Tr}_{R}\ket{\Psi}\bra{\Psi}=\sum\lambda_{i}\ket{u_{i}}\bra{u_{i}}_{L}$ 
where $L(R)$ stands for the left (right) trap in the double-well configuration 
and $\lambda_i$ are the eigenvalues of the Schmidt decomposition. 
The difference between the two largest non-degenerate Schmidt eigenvalues of the entanglement 
spectrum $\Delta\lambda=\lambda_{1}-\lambda_{2}$, is called the Schmidt gap, and it closes at 
the critical point in the TL \cite{DeChiara2012,Lepori2013}. 
% Moreover, in the Ising universality class, it has been shown that the Schmidt gap behaves as the magnetization, 
% and they share the same critical exponents \cite{DeChiara2012}. 

\begin{figure}[t!]
\centering
\epsfig{file=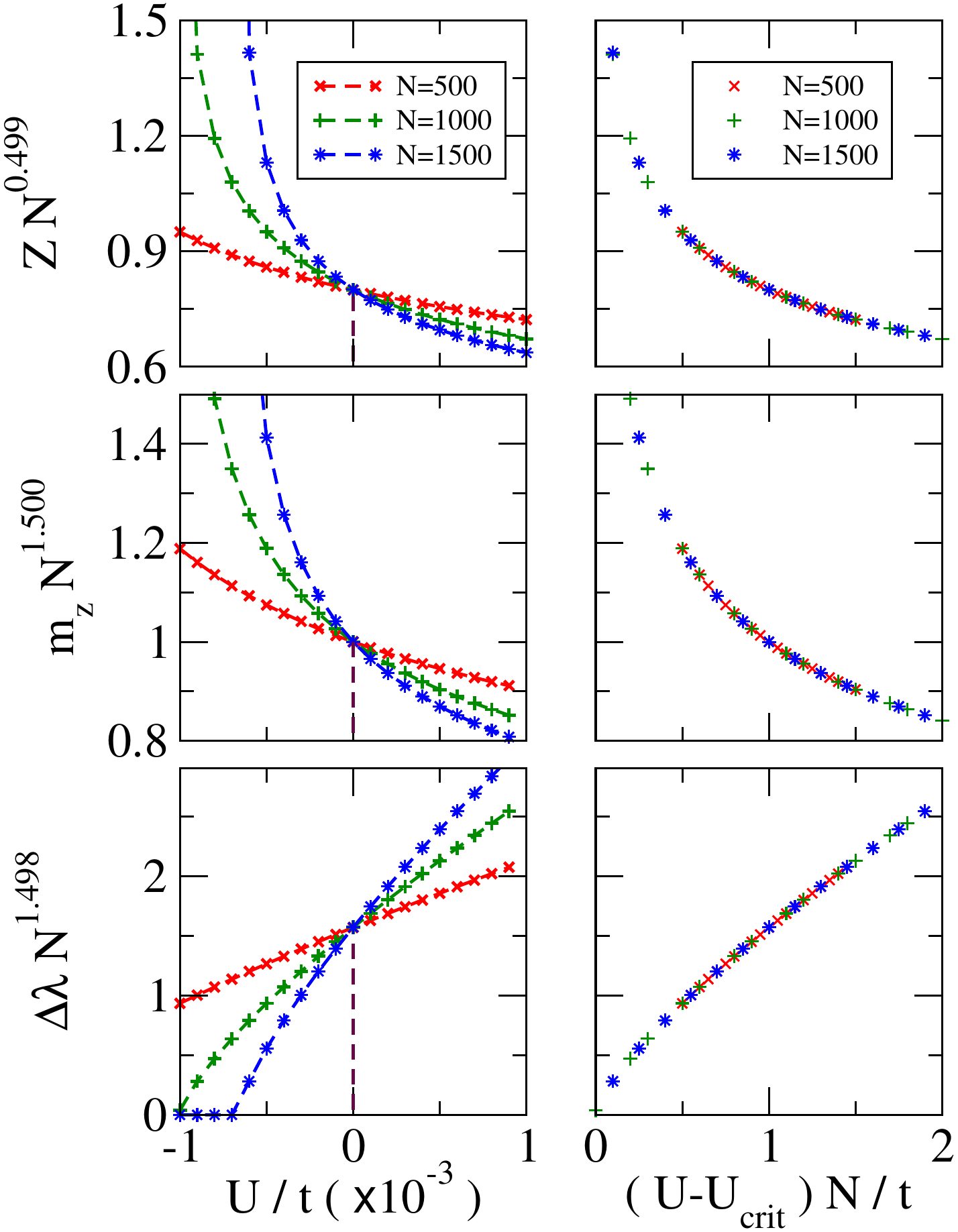, width=0.9\linewidth, clip=true}
\caption{Finite-size scaling behavior of the population imbalance (first row), the magnetization along 
the $z$-axis (second row) and the Schmidt gap (third row) in the double-well potential. The critical 
exponents obtained via this method are summarized in Table \ref{tab}.}
\label{Fig3}
\end{figure}

In spin chains, finite-size effects strongly modify the location of the critical point where a 
quantum phase transition occurs, while distorts the general properties of the transition \cite{Fisher1972}. 
In order to do that, we repeat our simulations for different number of atoms $N=500, 1000, 1500$ 
and look at the exponents that make all data to cross (critical point), and to collapse in order to obtain the critical exponents of observables, which scale as $\hat{O} \simeq N^{-\beta/\nu} f(|U-U_{crit}|N^{1/\nu})$, where $\nu$ is the mass gap exponent (associated to the correlation length divergence), and $\beta$ is the critical exponent of the corresponding operator $\hat{O}$. 
In Fig. \ref{Fig3} we display the scaling behavior of the population imbalance $Z$, the magnetization along the $z$-axis, 
and the Schmidt gap for the double-well case. Despite the fact that $\Delta\lambda$ is not an observable, all quantities exhibit 
scaling, which allows us to extract the critical exponents, which are summarized in Table \ref{tab}.

\begin{table}[h!]
 \begin{center}
      \begin{tabular}{| c || c | c | c |c |}
    \hline
  & $Z$ & $\Delta Z$ &$m_{z}$& $\Delta \lambda$  \\
    \hline
    \hline
    $\nu$ & 1.000 & 1.000 & 1.000 & 1.000  \\
    $\beta$ &  0.499 & 0.502 & 1.500 & 1.498\\
    \hline
    \end{tabular}
 \end{center}
 \caption{Critical exponents of the correlation length ($\nu$) and the scaling operator ($\beta$) obtained from 
 the scaling of the population imbalance ($Z$), its fluctuations ($\Delta Z$),  the magnetization 
 along the $z$-direction $m_{z}$, and the Schmidt gap $\Delta\lambda$ (obtained from the entanglement spectrum).}
\label{tab}
\end{table}

As expected, from the scaling of population imbalance, we obtain the mean-field exponents for $S_{z}$ and the mass gap, 
which converge to $\beta=1/2$ and $\nu=1$ in agreement with previous works \cite{Dusuel2004,Buonsante2012}. 
Interestingly, the critical exponents obtained from the scaling of the magnetization $m_z$ and the entanglement 
spectrum $\Delta\lambda$ coincide and result into $\beta=3/2$ and $\nu=1$ in agreement with \cite{Botet1983,Reslen2005}.

\section{Quantum phase transitions in the extended Bose-Hubbard Hamiltonian: universality}

We have seen in the previous section that the second order QPT that appears in 
the phase diagram of the double-well system, can be accurately described in terms 
of a correlation function inspired by a spin model like the Ising spin chain. The 
reinterpretation of the occupation number on each well as an effective length in 
the corresponding Hilbert space allows to provide scaling properties of some important 
observables like population imbalance, magnetization along a given axis or the Schmidt 
gap. In turn, the scaling properties of such observables lead to the a set of critical 
exponents that can be associated to the corresponding quantum phase transition. A 
natural question arises when the number of wells is increased:  Does the same analysis 
hold outside of the double-well configuration?

\begin{figure}[t!]
\centering
\epsfig{file=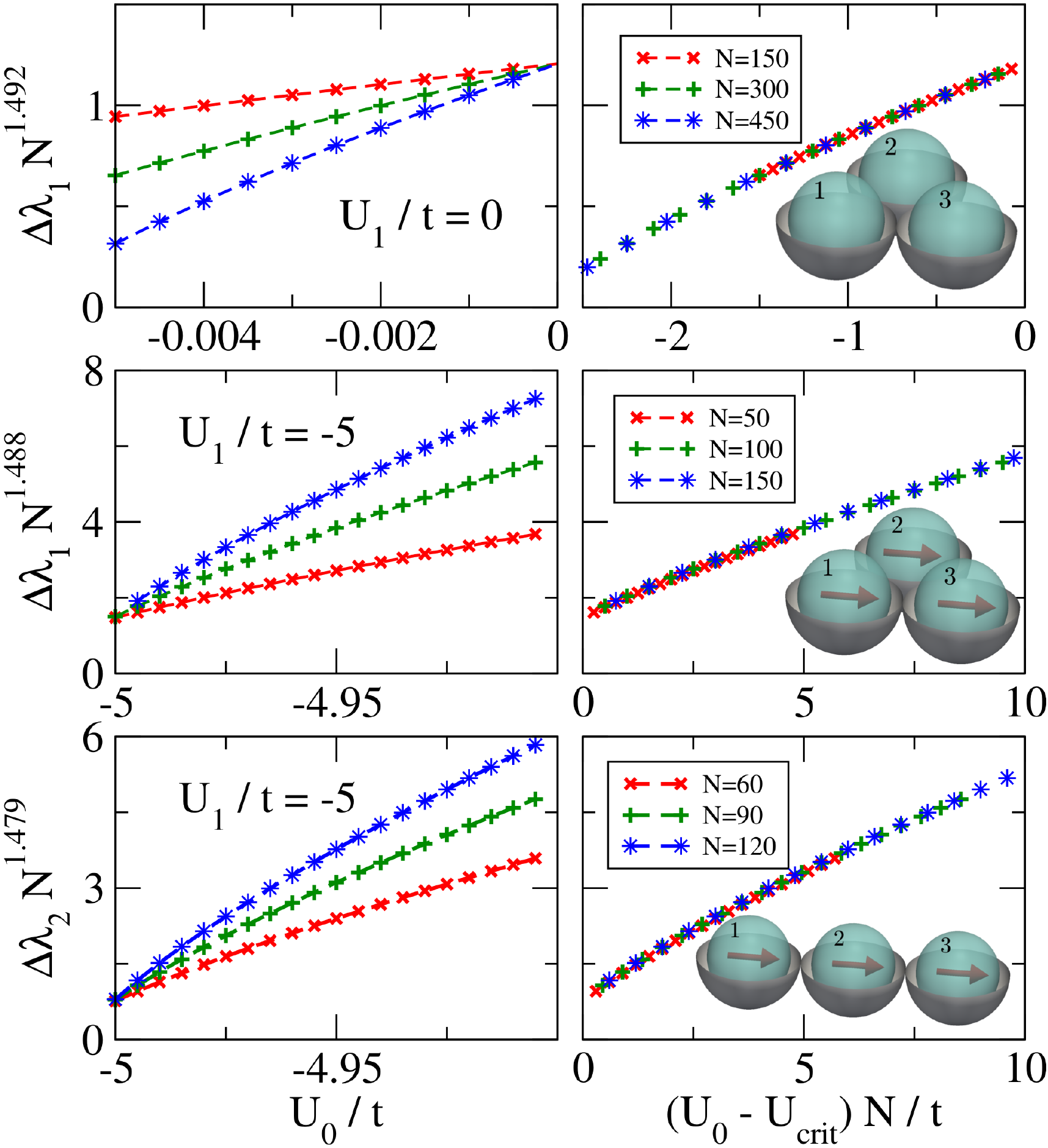, width=0.97\linewidth, clip=true}
\caption{Finite-size scaling of the Schmidt gap $\Delta\lambda_1$ (top and middle rows) and 
$\Delta\lambda_2$ (bottom row) for the three configurations displayed in the insets of 
the right panels. The critical exponents obtained are reported in the text. 
}
\label{Fig4}
\end{figure}

In this section, we focus on the study of the solutions of the extended Bose-Hubbard   
Hamiltonian given by (\ref{BHH}) for the triple-well configuration, including now the 
role of dipolar interactions. Depending on 
the geometry of the wells, and the polarization orientation, the system will display 
the anisotropy and the long range of the dipolar interaction. Three cases, which are 
particularly appealing for our aim, have been studied. The first one corresponds to 
the three sites constituting a triangular geometry with no dipolar interaction 
\cite{Dell'Anna2013} 
% 
% as depicted in the left column of Fig. 4. Such system crosses a second order quantum 
% phase transition for negatives values of the on-site interaction as reported in [27,28]. 
% The second system we analyze shares the same geometry but with dipolar interactions 
% now included as schematically indicated  by the arrows in the inset of the Fig.4. All 
% dipoles are oriented along the same direction.  Previous studies [29] have shown that 
% under such polarization starting from a configuration where the three traps have the 
% same population and fixing the value of the dipolar interactions, there is a quantum 
% phase transition to a final configuration in which one of the traps is empty while 
% the remaining two behave effectively as a double-well. Finally, we also analyzed the 
% triple-well configuration with all wells aligned along the ...... This last system 
% presents several quantum phase transitions depending on the ratio between the on-site 
% and dipolar interactions [30].  In particular, the system exhibits a QPT at negative 
% values of the on-site interactions and the system......
%
(see the inset of the top-right panel of Fig. \ref{Fig4}), which 
owns a QPT at negatives values of the on-site interaction \cite{Gallemi2015a}, 
analogous to the double-well case. The second one corresponds to the same triangular 
geometry holding dipolar atoms, with all the dipoles oriented along 
the direction defined by two of the sites, according to the inset of the middle-right panel 
of Fig. \ref{Fig4}. Under this polarization orientation, the system exhibits a QPT, where one of 
the sites, is empty, and the system behaves effectively as a double well \cite{Gallemi2013}. 
Finally, the third case concerns a three-well linear configuration with all sites aligned along the 
polarization direction of the dipoles \cite{Lahaye2010}, as shown in the inset of the 
bottom-right panel of Fig. \ref{Fig4}. The phase diagram of this system 
presents a QPT at negatives values of the on-site interaction, and the system transits 
from a state where all the atoms are in the central well, to another state where half of the 
atoms are in the central site and the other half macroscopically occupies one of the external wells, 
constituting a cat state.

% Going from the double-well to the triple-well configuration opens the question on how to extend 
% to the multipartite case some of the observables which are defined only for bipartite (double-well) 
% systems.  Here we focus on the Schmidt gap observable and define for each of the possible partitions 
% 1/23 and 2/13 (the partition 3/12 is equivalent to 2/13) the scaling properties of the it.  Our 
% results are summarized in Fig. 4 where we represent, for each of the analyzed systems, the scaling 
% properties of the Schmidt gap. On the left column, we display the crossing of the Schmidt gap 
% giving raise the precise localization of the critical point where the QPT occurs. On the right 
% column, the collapse of the data provides the critical exponent $\mu$.  
In the double-well case, we have analyzed the scaling properties of some observables, like 
the imbalance, its fluctuations and the magnetization. For the triple-well case, the 
definition of the equivalent observables is not so straightforward. Nevertheless, the Schmidt 
gap can be computed in the way commented before, and we can take profit from its scaling 
properties. In three-well systems, however, there are more than a single partition of the 
system (remember that the Schmidt decomposition only works for bipartite systems). In particular, 
the last two systems presented in the previous paragraph, have two independent partitions 
that lead to two different sets of Schmidt gaps. Following the notation of the labels of 
the sites in the insets of Fig. \ref{Fig4}, $\Delta\lambda_1$ corresponds to the $1/23$ 
partition ($1/23$ must be read as the partition of site $1$ with respect to the subsystem 
constituted by sites $2$ and $3$), whereas, $\Delta\lambda_2$ corresponds to the $2/13$ partition.

The top (middle) panels of Fig. \ref{Fig4} represent the scaling of $\Delta \lambda_1$ in the 
triangular non-dipolar (dipolar) case for $N=150,300,450$ ($N=50,100,150$), in the transitions 
commented above at $U_1=0$ ($U_1/t=-5$), where $U_1$ accounts for the dipole-dipole interaction 
strength, see Ref. \cite{Gallemi2013} for details. The bottom panels show the scaling of $\Delta\lambda_2$ 
for the aligned case for $N=60,90,120$. In all three cases, the critical exponents are: 
$\beta=3/2$ and $\nu=1$, which are the same that the ones obtained in the scaling for the double well. 
As a consequence, they share the universality class with the double-well transition \cite{Gallemi2013}. This 
property becomes evident for the triangular case with dipolar atoms, where across the QPT analyzed 
previously, the system behaves as an effective double-well system, with one of the wells completely 
empty.

\section{Summary and conclusions}

We have analyzed in this paper the QPTs appearing for mesoscopic BECs trapped in double and in a triple-well 
configurations. For the former case, the system can be mapped 
to the infinite-range Ising model (Lipkin-Meshkov-Glick), and we can define the concept 
of length and translational invariance in the Hilbert space. This allows us to demonstrate that 
%mean-field ....
%We have analyzed the QPTs of some few-site Bose-Hubbard hamiltonians describing mesoscopic Bose-Einstein 
%condensates. We have demonstrated that in some cases, they can be mapped onto spin models, and in particular, 
%for the double-well case, we have explicitly performed the mapping onto the Lipkin-Meshkov-Glick hamiltonian. 
%We have incorporated in this particular setup, the concept of length and translational invariance in the 
%Hilbert space, which allows to demonstrate that 
mean-field QPTs can be also associated to the divergence of a correlation length. 
%the critical exponents so obtained fit those of the LMG model. 
% 
% 
Finally, we have generalized our study to the triple-well extended Bose-Hubbard Hamiltonian, which includes 
also dipolar interactions. We have shown that there exists few QPTs that belong to the same universality 
class of the double-well (Lipkin-Meshkov-Glick) transition, as reflected by the fact that all the QPTs 
share the same critical exponents obtained by the scaling properties of the corresponding Schmidt gap. 
%There are also other types of QPTs that are not at all related to the LMG model and whose critical 
% exponents are completely different from the mean-field Ising model.  These facts strongly support .....
%
%By using finite-size scaling of some observables, like the population imbalance or 
%the magnetization, with the number of atoms, we find that some critical exponents of the QPT fit those 
%of the Lipkin-Meshkov-Glick transition. We have also analyzed the scaling of the Schmidt gap in this system, which 
%exhibits the same critical exponents of the magnetization, as demonstrated in Ref. \cite{DeChiara2012} for the 
%Ising model. 
%
%We have extended our hamiltonian to the triple-well case, including dipolar atoms. We have shown that there 
%exist few QPTs in these extended hamiltonians whose analysis of the Schmidt gap displays a scaling with the 
%number of atoms with the same critical exponents of the double-well case and the Lipkin-Meshkov-Glick model. 
%Hence, all these QPTs fall in the same universality class \cite{Mussardo2010}. 
These facts strongly support 
the suitability of mesoscopic Bose-Einstein condensates as quantum simulators of condensed matter physics.

\section*{Acknowledgments}

We acknowledge financial support from the Spanish MINECO (FIS2014-52285-C2-1-P and FIS2014-4062-P) and the 
European Regional development Fund, Generalitat de Catalunya Grant No. SGR2014-401 and SGR2014-946. A. G. 
is supported by Generalitat de Catalunya Grant FI-DGR 2014 and Spanish MECD fellowship FPU13/02106. We 
thank useful discussions with G. De Chiara, S. Campbell and M. Moreno-Cardoner.

\bibliography{bibtex.bib}

\end{document}